\documentstyle[11pt,newpasp,twoside,epsf,graphics,amssymb]{article}
\def\edcomment#1{\iffalse\marginpar{\raggedright\sl#1\/}\else\relax\fi}
\reversemarginpar

\begin{document}
\title{Properties of Low-Redshift Damped Lyman Alpha Galaxies}
\author{Daniel B. Nestor, Sandhya M. Rao, David A. Turnshek}
\affil{Department of Physics and Astronomy, University of Pittsburgh, 
Pittsburgh, PA 15260}
\author{Eric Monier}
\affil{Department of Astronomy, Ohio State University, Columbus, OH 43210}
\author{Wendy Lane}
\affil{Naval Research Laboratory, 4555 Overlook Ave. S.W., Washington DC 20375}
\author{Jacqueline Bergeron}
\affil{ESO, Karl-Schwarzschild-Strasse 2, Garching bei Munchen D-85748,
Germany}


\begin{abstract}
Images of five QSO fields containing six damped Ly$\alpha$ (DLA) systems
at redshifts $0.09<z<0.53$ are presented. Identifications for the DLA
galaxies giving rise to the DLA systems are made. The observed and modeled
characteristics of the DLA galaxies are discussed.  The DLA galaxies
have impact parameters ranging from $<4$ kpc to $\approx 34$ kpc and
luminosities in the range $\approx 0.03$L$^*$ to $\approx 1.3$L$^*$.
Their morphologies include amorphous low surface brightness systems,
a probable dwarf spiral, and luminous spirals.  \end{abstract}

\section{Introduction}
Since damped Ly$\alpha$ systems (DLAs) identified in QSO 
absorption-line surveys trace the bulk of the observable neutral gas in the
Universe from high to low redshift ($4>z>0$), they are important
probes of galaxy formation and evolution. In principle, their 
observed cosmic evolution is the net result of the global processes
which give rise to conversion between the HII, HI, and H$_{2}$ 
phases. Thus, while galaxy populations
selected in optical/IR surveys can be used to study the star
formation history of the Universe, DLA galaxies identified through
follow-up imaging work, i.e., by virtue of their gas cross-sections,
reveal the HI gas production and consumption history of the Universe.
At present, only at low redshift can we easily study and compare
these populations.  With the discovery of a significant number
of low-redshift ($z<1.65$) DLAs (Rao \& Turnshek 2000), we are
now in a position  to comprehensively study the properties of
the DLA galaxies that give rise to them.  Le Brun et al. (1997)
imaged a sample of six low-redshift DLA fields with HST, and found
that DLA galaxies span a range of luminosities and morphologies.
Here, we present ground-based imaging results of five QSO fields
with six DLAs and confirm their results. Overall, the DLA galaxy
population at low-redshift contains a significant number of dwarf
and/or low surface brightness (LSB) galaxies.  Results on two DLAs
in the Q0738+313 field were presented in Turnshek et al. (2001),
and we summarize them here.  In Turnshek, Rao, \& Nestor (2001, these
proceedings), we compare the properties of low-redshift DLA galaxies
with the population  of gas-rich galaxies in the local Universe.
Throughout this paper, we use H$_0=65$ km s$^{-1}$ Mpc$^{-1}$,
q$_0=0.5$, and $\Lambda=0$.

\section{Observations}
Optical and infrared images of the five QSO fields discussed here
were obtained during the period between November 1997 and September
2000.  The optical images were obtained at the MDM Observatory 2.4m
Hiltner Telescope on Kitt Peak using the $1024\times1024$ Templeton
CCD (0.285$\arcsec$ pixel$^{-1}$) and by the queue observing team
using a $1024\times1024$ Tektronix CCD (0.195$\arcsec$ pixel$^{-1}$)
on the 3.5m WIYN Telescope on Kitt Peak.  The infrared images
were obtained at the 3.0m NASA IRTF on Mauna Kea using NSFCAM,
a $256\times256$ InSb detector array (0.30$\arcsec$ pixel$^{-1}$)
and the 3.58m ESO NTT on La Silla using a $1024\times1024$ HgCdTe
detector array (0.292$\arcsec$ pixel$^{-1}$).  The images were
reduced using the usual procedures and standard star observations
were used to calibrate the photometry.

\section{SED and Isophotal Template Fits}
In order to explore the properties of three of the identified DLA
galaxies, spectral energy distribution (SED) templates were fitted to the
available multiband photometric data. Using the GISSEL99 (Bruzual and
Charlot) library of SEDs, stellar populations corresponding to eleven
different ages (spanning 0.001 to 12.0 Gyrs) derived from a Salpeter 
IMF were combined in all possible permutations of two different ages. Each
permutation was then combined with eleven  different weighting
schemes. The resulting 605 templates were subjected to a Calzetti
(2000) reddening law with eleven possibilities for the amount of dust
(spanning $0.0 \le$ A$_{V} \le 3.6$).  Each of the eleven sets (corresponding
to a given amount of dust) was evaluated with a principal component
analysis routine in order to extract eigen-spectra for each set. The
template built from the eigen-spectra giving the smallest chi-square
when compared to the DLA galaxy photometry was then mapped back to the
original set of templates to determine the ages, reddenings, and stellar
populations that best fit the DLA galaxy photometry. The results are
summarized in \S4 and given in Table 1 and Figure 1a.

We also investigated the radial light profiles of the DLA galaxies.
Exponential and r$^{1/4}$ profiles convolved with a
Gaussian to model the effects of seeing were compared to the observed
profiles.  It was not possible to fit isophots for the three DLA galaxies
that were classified as LSB/amorphous and the results for one of the
luminous (probable) spirals were inconclusive. The results for two
of the identified DLA galaxies are summarized 
in \S4 and presented in Table 1 and Figure 1b.

\begin{figure}[h]
\plottwo{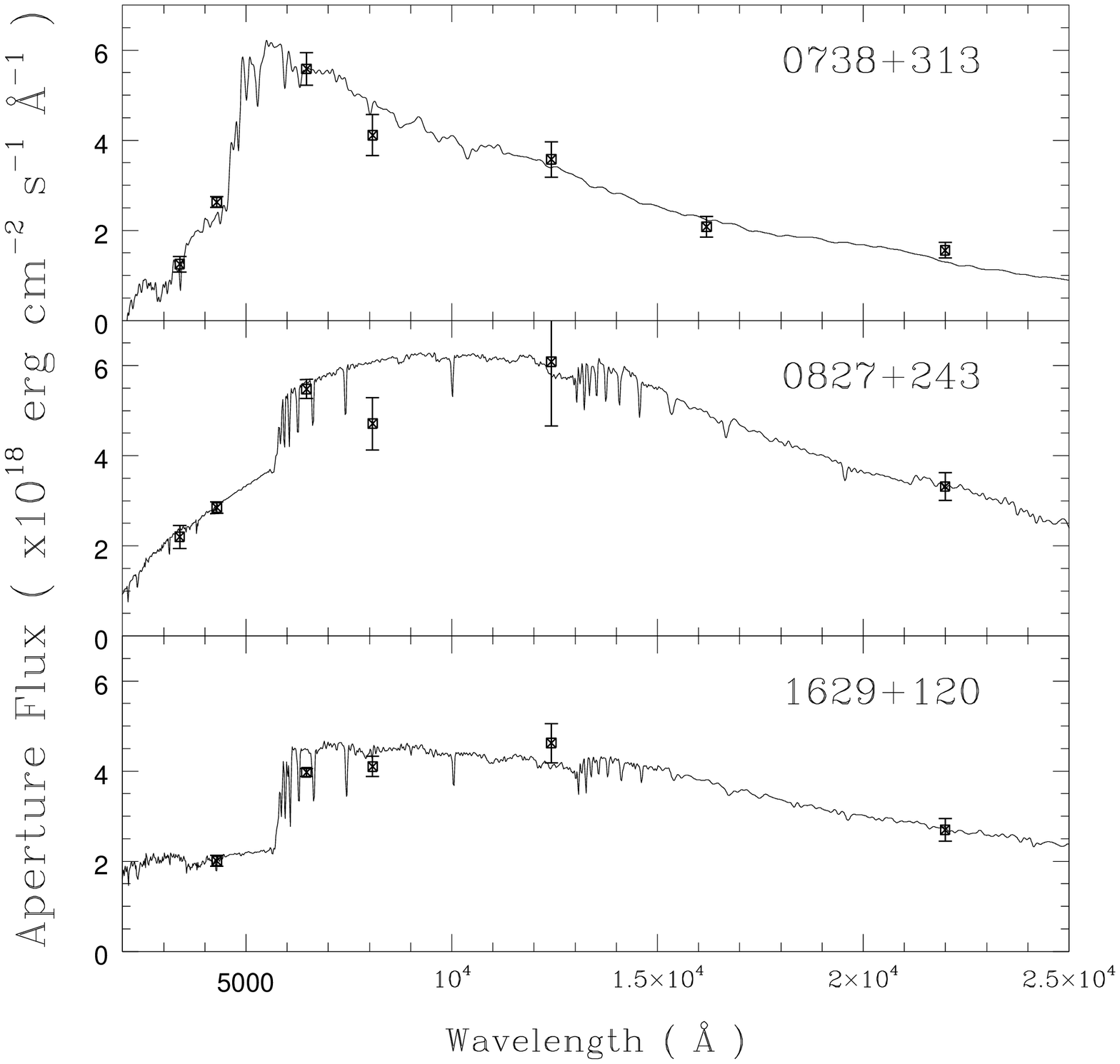}{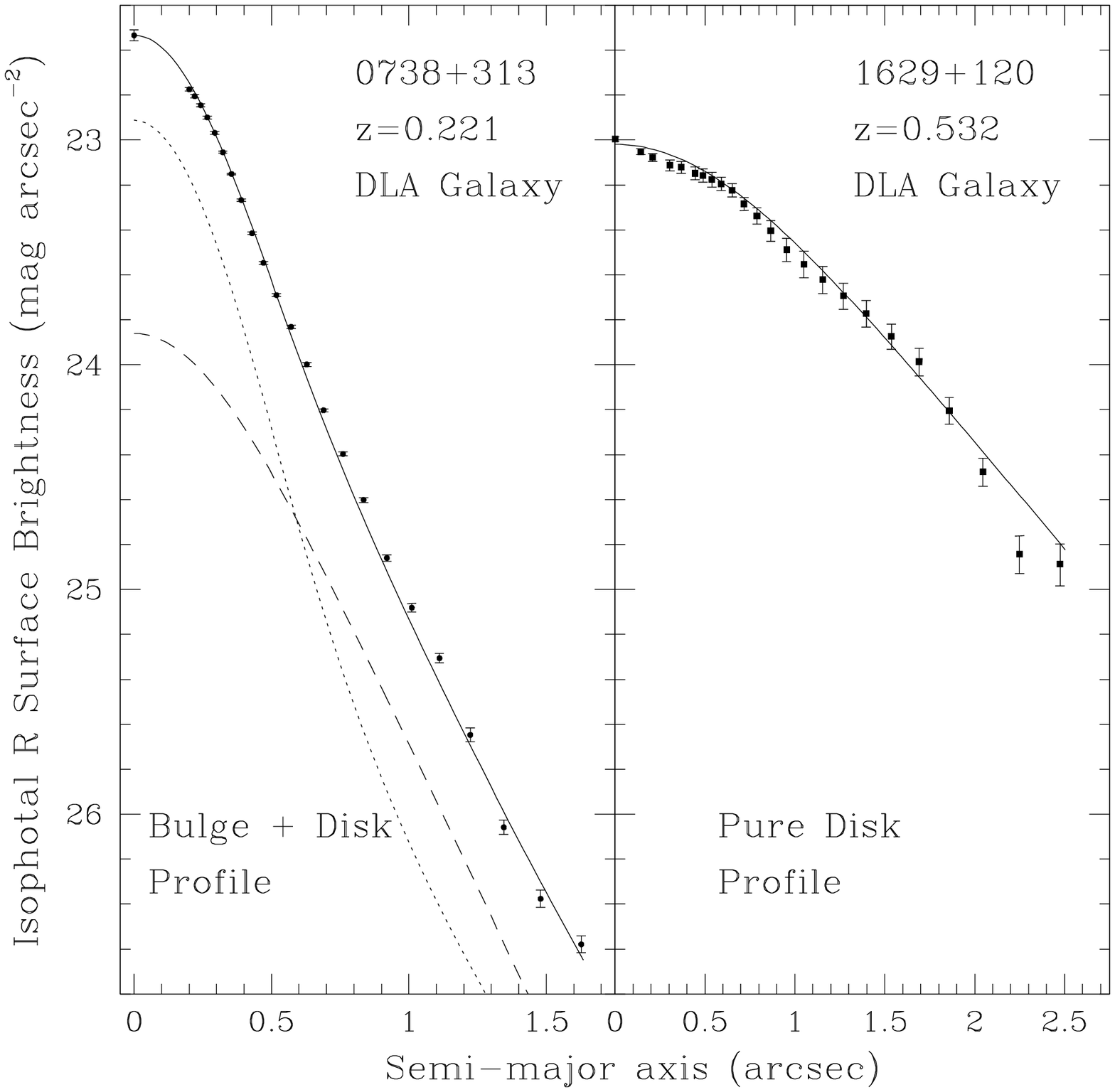}
\caption{\footnotesize
The left three-panel figure shows best-fit two-burst
templates for three DLA galaxies.  See Table 1 for details of each
template. Data points are UBRIJHK photometric measurements from the
MDM, WIYN, and IRTF telescopes.  The right two-panel figure shows
model fits to the R-band radial light profiles of two DLA galaxies.
In the left panel is the fit for the $z=0.221$ DLA galaxy towards
Q0738+313. The interior isophots are dominated by an r$^{1/4}$
(bulge) profile while the outer isophots are dominated by an
exponential (disk) profile.  In the right panel is the fit for the
$z=0.532$ DLA galaxy towards Q1629+120.  A purely exponential (disk)
profile is a good fit to these data.} \end{figure}

\section{DLA Galaxies}
Figures $2-6$ show the five DLA QSO fields that contain six DLA
galaxies.  Three DLA galaxies have a confirmed slit redshift and
are labeled in the corresponding figure.  Identifications are also
made for the other three DLAs.  It should be kept in mind, however,
that the identification of a DLA galaxy is never fully certain.
While one can be fairly confident with an identification when
the galaxy appears to be isolated and at low impact parameter to
the sightline, it is always possible  that another faint galaxy
at low impact parameter might be hidden in the PSF of the QSO.
See Turnshek, Rao, \& Nestor (2001, these proceedings) for further
discussion.  We consider the identifications and properties of the
six DLA galaxies below.

\begin{figure}[t]
\plotfiddle{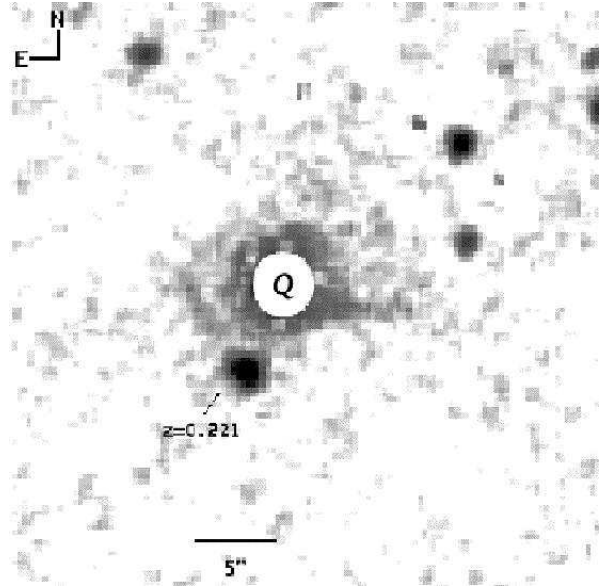}{8.5cm}{0}{40}{40}{-110}{-10}
\caption{\footnotesize
The IRTF K band image of the Q0738+313 field.  The light at the
position of the QSO (labeled Q) has been subtracted.  The DLAs are at
redshifts $z=0.091$ and $z=0.221$.  On the image, 5 arcsec corresponds
to 9 kpc at $z=0.091$ and 17.5 kpc at $z=0.221$.  The $z=0.091$ DLA is
assumed to be the LSB galaxy at low impact parameter surrounding the
sightline (Turnshek et al. 2001).} \end{figure}

\begin{figure}
\plotfiddle{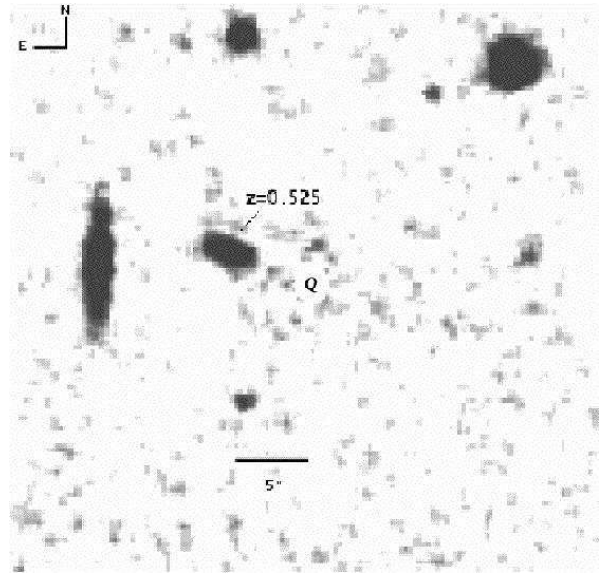}{6.5cm}{0}{40}{40}{-110}{-10}
\caption{\footnotesize
IRTF K band image of the Q0827+243 field.  The light at the
position of the QSO (labeled Q) has been subtracted.  The DLA  redshift
is $z=0.525$. On the image, 5 arcsec corresponds to 28 kpc at the DLA
redshift.} \end{figure}

\begin{figure}
\plotfiddle{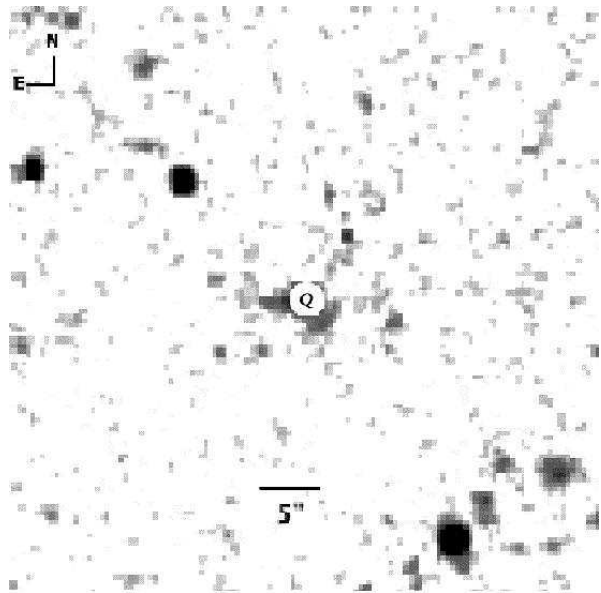}{6.5cm}{0}{40}{40}{-110}{-10}
\caption{\footnotesize
The NTT J band image of the Q0952+179 field containing a
$z=0.239$ DLA absorber.  The light  at the position of the QSO (labeled Q)
has been subtracted.  The LSB structure to the east and southwest of the
sightline is identified as the absorber given the absence of brighter
objects near the sightline. On the image 5 arcsec corresponds to
18.5 kpc at $z=0.239$} \end{figure}

\begin{figure}
\plotfiddle{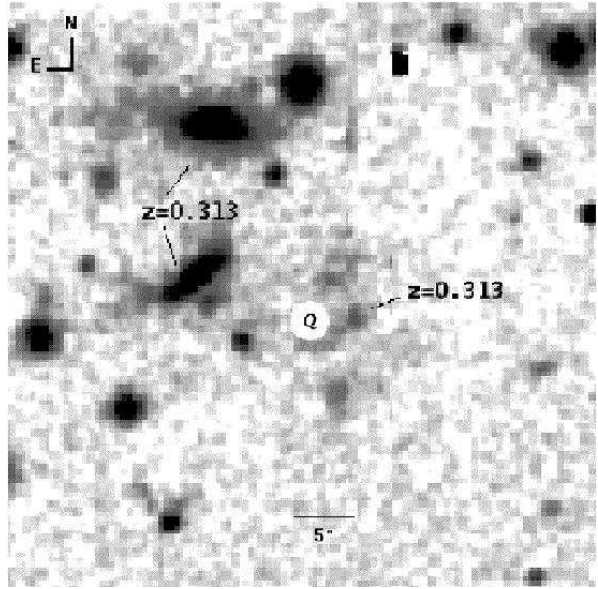}{6.5cm}{0}{40}{40}{-110}{-10}
\caption{\footnotesize
The NTT J band image of the Q1127$-$145 field containing a
$z=0.313$ DLA absorber.  The light  at the position of the QSO (labeled Q)
has been subtracted.  The LSB structure extending north-south and to the 
west of the
QSO is identified as the absorber.  On the image, 5 arcsec corresponds
to 22 kpc at the DLA redshift.} \end{figure}

\begin{figure}
\plotfiddle{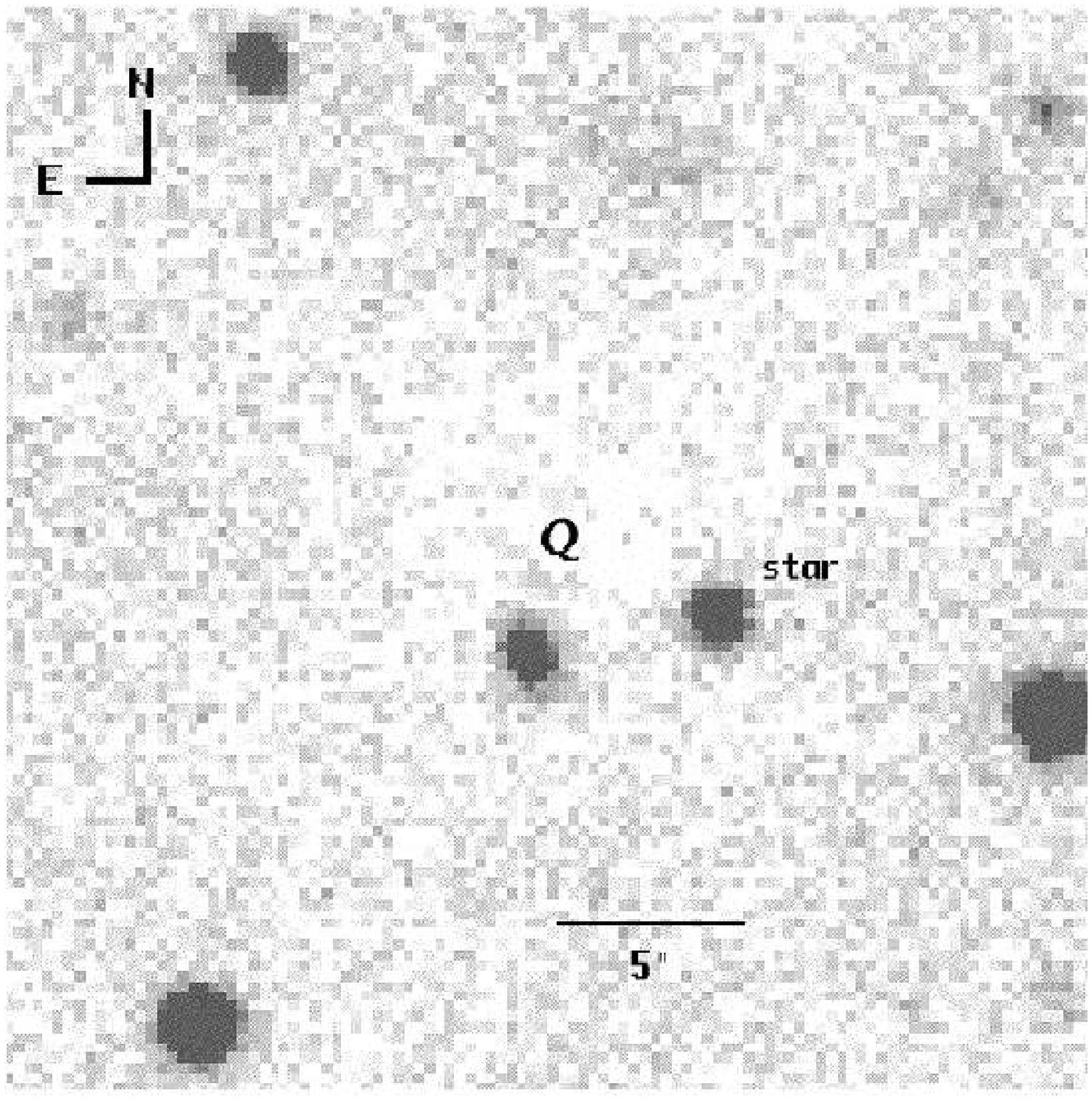}{6.5cm}{0}{40}{40}{-110}{-10}
\caption{\footnotesize
IRTF K band image of the Q1629+120 field containing a $z=0.532$
DLA absorber. The light at the position of the QSO (labeled Q) has been
subtracted.  The luminous galaxy to the south of the QSO is identified
as the DLA galaxy.  On the image, 5 arcsec corresponds to 28 kpc at the
DLA redshift.} \end{figure}

The sightline towards {\bf Q0738+313} contains two low-redshift
DLAs at $z=0.091$ and $z=0.221$.  The $z=0.091$ absorber is
identified with the LSB galaxy at low impact parameter (see
Figure 2).  A very red (R$-$K $>5.5$) ``arm''-like feature can be
seen in the infrared images to the east of the QSO, suggestive of
spiral structure.  The ``jet''-like feature to the west, however,
is bluer (R$-$K $\approx3.2$) which suggests the possibility of
an irregular/interacting system.  The total observed extent of
the system is $\approx12-16$ kpc and the upper-limit on its
luminosity is $\approx0.13$L$^{*}$.  The $z=0.221$ absorber
is a an early type dwarf (0.1L$^{*}$) at an impact parameter of
20 kpc.  Its colors are consistent with star formation models
suggesting its formation epoch was less than a few Gyr ago, i.e.,
$z_f\approx0.3-0.9$ (Figure 1a).  Its isophotal light distribution
indicates a combination of r$^{1/4}$ bulge-like and exponential
disk-like components (Figure 1b).  These systems are discussed in
more detail in Turnshek et al. (2001).

The sightline towards {\bf Q0827+243} contains a DLA at $z=0.525$.
The luminous ($\approx1.3$L$^{*}$) galaxy 6 arcsec,
or 34 kpc, to the east of the sightline is identified as the DLA galaxy
(see Figure 3).
Although isophotal profile fitting was inconclusive, the
colors are consistent with star formation models suggesting recent star
formation and the presence of significant dust (Figure 1a).  The galaxy
is likely a late-type object and possibly a spiral, 
although this remains to be confirmed.

The sightline towards {\bf Q0952+179} contains a DLA at $z=0.239$.
The structure seen to the east and southwest of the QSO is the only
evidence for an absorber and therefore is identified as the DLA galaxy
(see Figure 4).  The galaxy is inferred to be LSB, but its faintness and
small impact parameter ($<7$ kpc) make it difficult to tell whether it
is a single LSB galaxy or a patchy/irregular system.  A PSF subtraction
of the light from the QSO leads to the estimated lower-limit on
its luminosity of $\approx0.03$L$^{*}$.

The sightline towards {\bf Q1127$-$145} contains a DLA at $z=0.313$.
This DLA is associated with the  north-south elongated patchy/irregular
LSB structure $\approx3.5$ arcsec to the west of the sightline
(see Figure 5).  A PSF subtraction of the light from the QSO gives a
luminosity of $\approx0.05$L$^{*}$ for the DLA galaxy. Two other bright
objects in the field are at the absorption redshift as well; evidently
there is a galaxy group or cluster at this redshift.  The LSB structure
is identified as the DLA absorber due to its smaller impact parameter
($<10$ kpc as opposed to $>36$ kpc for the luminous galaxies).

The sightline towards {\bf Q1629+120} contains a DLA at $z=0.532$.
The luminous ($\approx 1.1$L$^{*}$) galaxy 3 arcsec, or 17 kpc,
to the south of the sightline is the only resolved object in the field
(see Figure 6) leading to the identification of this object as the DLA
galaxy. A pure exponential law fits the isophotal light profile of this
object well, and SED template fits to its colors imply recent  star
formation and the presence of dust (Figures 1a and 1b). Taken together,
the observed and modeled characteristics strongly imply that this DLA
galaxy is a luminous late-type spiral.

\begin{center}
\begin{table}
\scalebox{.8}{
\begin{tabular}{|ccccccc|}
\multicolumn{7}{c}{\LARGE{\bf Table 1: DLA Galaxy Characteristics}}\\ 
\hline 
\multicolumn{7}{|c|}{}\\
{\bf QSO} & {\bf 0738+313} & {\bf 0738+313} & {\bf 0952+179} & {\bf 1127$-$145} & {\bf 0827+243} & {\bf 1629+120}\\
\multicolumn{7}{|c|}{}\\ 
\hline 
\hline 
\multicolumn{7}{|c|}{}\\ 
{\bf z$_{\rm abs}$} & 0.091 & 0.221 & 0.239 &0.313 & 0.525 & 0.532 \\ 
& & & & & & \\ 
{\bf N$_{\rm\bf HI}$} & $1.5\times10^{21}$ & $7.9\times10^{20}$ & 2.1$\times10^{21}$  & 5.1$\times10^{21}$ &
$2.0\times10^{20}$ &  $2.8\times10^{20}$ \\   
(cm$^{-2}$) & & & & & & \\ 
& & & & & & \\
{\bf L\tablenotemark{\bf a}}& $\approx$ 0.08 L$^{*}$ & $\approx$ 0.10
L$^{*}$ & $\gtrsim$ 0.03 L$^{*}$ & $\approx$ 0.05 L$^{*}$ & $\approx$
1.3 L$^{*}$ &  $\approx$ 1.1 L$^{*}$ \\ 
& & & & & & \\ 
{\bf b\tablenotemark{\bf b}} & $<4$ & 20 & $<7$  & $<10$ & 34 &  17 \\
(kpc) &  & & & & & \\ 
& & & & & & \\ 
{\bf Morphology} & LSB & compact & patchy/ & patchy/ & possible  & spiral \\ 
                 &     &  dwarf  & irr LSB & irr LSB &   spiral  &      \\ 
& & & & & & \\ 
\hline   
\multicolumn{7}{|c|}{\bfseries }\\
\multicolumn{7}{|c|}{\bfseries Best-Fit Model Template}\\
\hline 
\multicolumn{7}{|c|}{}\\ 
{\bf Burst Ages\tablenotemark{\bf c}} &$\cdots$  & 80\% 0.6 Gyr& $\cdots$ & $\cdots$ & 74\% 0.001 Gyr&  90\%
0.2 Gyr \\ & $\cdots$  & 20\% 4.0 Gyr & $\cdots$&  $\cdots$ & 26\%
0.01 Gyr & 10\% 0.001 Gyr \\ 
& & & & & & \\ 
{\bf A$_{\rm\bf V}$} & $\cdots$  &   0.4 & $\cdots$&  $\cdots$ & 3.6 & 2.0  \\ 
& & & & & & \\ 
{\bf Isophotal Fit} & $\cdots$ & disk $+$ bulge & $\cdots$& $\cdots$ & $\cdots$ & disk \\  
& & & & & & \\ 
\hline
\end{tabular}
\tablenotetext{a}{Based on infrared photometric data.}
\tablenotetext{b}{b $\equiv$ impact parameter.}
\tablenotetext{c}{Fraction by burst-mass, best-fit two-burst model.}
}
\end{table}
\end{center}

\section{Summary}
The DLA galaxies presented here span a wide range of observed
characteristics (see Table 1). The neutral hydrogen column densities range
from the minimum value for a classical DLA system ($N_{HI}=2\times10^{20}$
atoms cm$^{-2}$) up to $5.1\times10^{21}$ atoms cm$^{-2}$. Similarly, the
luminosities range from $\approx0.03$L$^{*}$ to 1.3L$^{*}$. The largest
impact parameter in this small sample is 34 kpc, while three of the
systems have impact parameters less than 10 kpc.  The morphologies
likewise vary. One of the luminous galaxies is a spiral, while the
other is a possible spiral; three are underluminous LSB galaxies with
patchy/irregular structure; and one is a compact dwarf galaxy. Thus, the range
in DLA galaxy properties seen here is consistent with previous findings that 
they are drawn from a variety of galaxy types (Le Brun et al.
1997; Steidel et al. 1997; Rao \& Turnshek 1998; Turnshek et al. 2001).

This material is based upon work supported by the National Science 
Foundation under Grant No. 9970873.

\end{document}